\documentclass[aps,prl,a4paper,superscriptaddress,twocolumn,showpacs,amsmath,amssymb]{revtex4} 
\usepackage{graphicx,epsfig}
\usepackage[usenames]{color}

\usepackage{float}  

\allowdisplaybreaks
\begin{document}
\newcommand{\rl}{\hat{\rho}_{\text{loc}}}
\newcommand{\rs}{\hat{\rho}_{\mathcal{S}}}
\newcommand{\ms}{\mathcal{S}}
\newcommand{\mb}{\mathcal{B}}
\newcommand{\mU}{\mathcal{U}}
\newcommand{\hl}{\hat{\mathcal{H}}^{\text{loc}}}
\newcommand{\el}{\hat{e}_{\text{loc}}}
\newcommand{\fl}{\hat{\mathcal{H}}_{\text{loc}}+\sum_k\mu_k\hat{n}_k}
\newcommand{\nl}{e^{-\left(\beta_{\mathcal{S}}\hat{\mathcal{H}}_{\mathcal{S}}+\sum_{k}\mu_{\mathcal{S}k}\hat{n}_{\mathcal{S}k}\right)}}
\newcommand{\zl}{\Tr\!\left[ e^{-\left(\beta_{\mathcal{S}}\hat{\mathcal{H}}_{\mathcal{S}}+\sum_{k}\mu_{\mathcal{S}k}\hat{n}_{\mathcal{S}k}\right)} \right]}
\newcommand{\E}{\mathcal{E}}
\newcommand{\Lag}{\mathcal{L}}
\newcommand{\M}{\mathcal{M}}
\newcommand{\N}{\mathcal{N}}
\newcommand{\U}{\mathcal{U}}
\newcommand{\R}{\mathcal{R}}
\newcommand{\F}{\mathcal{F}}
\newcommand{\V}{\mathcal{V}}
\newcommand{\C}{\mathcal{C}}
\newcommand{\I}{\mathcal{I}}
\newcommand{\s}{\sigma}
\newcommand{\up}{\uparrow}
\newcommand{\dw}{\downarrow}
\newcommand{\h}{\hat{\mathcal{H}}}
\newcommand{\himp}{\hat{h}}
\newcommand{\g}{\mathcal{G}^{-1}_0}
\newcommand{\D}{\mathcal{D}}
\newcommand{\A}{\mathcal{A}}
\newcommand{\projs}{\hat{\mathcal{S}}_d}
\newcommand{\proj}{\hat{\mathcal{P}}_d}
\newcommand{\K}{\textbf{k}}
\newcommand{\Q}{\textbf{q}}
\newcommand{\T}{\tau_{\ast}}
\newcommand{\io}{i\omega_n}
\newcommand{\eps}{\varepsilon}
\newcommand{\+}{\dag}
\newcommand{\su}{\uparrow}
\newcommand{\giu}{\downarrow}
\newcommand{\0}[1]{\textbf{#1}}
\newcommand{\ca}{c^{\phantom{\dagger}}}
\newcommand{\cc}{c^\dagger}
\newcommand{\aaa}{a^{\phantom{\dagger}}}
\newcommand{\aac}{a^\dagger}
\newcommand{\bba}{b^{\phantom{\dagger}}}
\newcommand{\bbc}{b^\dagger}
\newcommand{\da}{d^{\phantom{\dagger}}}
\newcommand{\dc}{d^\dagger}
\newcommand{\fa}{f^{\phantom{\dagger}}}
\newcommand{\fc}{f^\dagger}
\newcommand{\ha}{h^{\phantom{\dagger}}}
\newcommand{\hc}{h^\dagger}
\newcommand{\be}{\begin{equation}}
\newcommand{\ee}{\end{equation}}
\newcommand{\bea}{\begin{eqnarray}}
\newcommand{\eea}{\end{eqnarray}}
\newcommand{\ba}{\begin{eqnarray*}}
\newcommand{\ea}{\end{eqnarray*}}
\newcommand{\dagga}{{\phantom{\dagger}}}
\newcommand{\bR}{\mathbf{R}}
\newcommand{\bQ}{\mathbf{Q}}
\newcommand{\bq}{\mathbf{q}}
\newcommand{\bqp}{\mathbf{q'}}
\newcommand{\bk}{\mathbf{k}}
\newcommand{\bh}{\mathbf{h}}
\newcommand{\bkp}{\mathbf{k'}}
\newcommand{\bp}{\mathbf{p}}
\newcommand{\bL}{\mathbf{L}}
\newcommand{\bRp}{\mathbf{R'}}
\newcommand{\bx}{\mathbf{x}}
\newcommand{\by}{\mathbf{y}}
\newcommand{\bz}{\mathbf{z}}
\newcommand{\br}{\mathbf{r}}
\newcommand{\Ima}{{\Im m}}
\newcommand{\Rea}{{\Re e}}
\newcommand{\Pj}[2]{|#1\rangle\langle #2|}
\newcommand{\ket}[1]{\vert#1\rangle}
\newcommand{\bra}[1]{\langle#1\vert}
\newcommand{\setof}[1]{\left\{#1\right\}}
\newcommand{\fract}[2]{\frac{\displaystyle #1}{\displaystyle #2}}
\newcommand{\Av}[2]{\langle #1|\,#2\,|#1\rangle}
\newcommand{\av}[1]{\langle #1 \rangle}
\newcommand{\Mel}[3]{\langle #1|#2\,|#3\rangle}
\newcommand{\Avs}[1]{\langle \,#1\,\rangle_0}
\newcommand{\eqn}[1]{(\ref{#1})}
\newcommand{\Tr}{\mathrm{Tr}}

\newcommand{\Vb}{\bar{\mathcal{V}}}
\newcommand{\Vd}{\Delta\mathcal{V}}
\def\P{P_{02}}
\newcommand{\Pb}{\bar{P}_{02}}
\newcommand{\Pd}{\Delta P_{02}}
\def\t{\theta_{02}}
\newcommand{\tb}{\bar{\theta}_{02}}
\newcommand{\td}{\Delta \theta_{02}}
\newcommand{\Rb}{\bar{R}}
\newcommand{\Rd}{\Delta R}

\title{Principle of Maximum Entanglement Entropy
and Local Physics of \\ Correlated many-body Electron-Systems}
\author{Nicola Lanat\`a}
\affiliation{Department of Physics and Astronomy, Rutgers University, Piscataway, New Jersey 08856-8019, USA} 
\author{Hugo U. R. Strand}
\affiliation{Department of Physics, University of Gothenburg, SE-412 96 Gothenburg, Sweden}
\affiliation{Department of Physics, University of Fribourg, CH-1700 Fribourg, Switzerland}
\author{Yongxin Yao}
\affiliation{Ames Laboratory-U.S. DOE and Department of Physics and Astronomy, Iowa State 
University, Ames, Iowa IA 50011, USA}
\author{Gabriel Kotliar}
\affiliation{Department of Physics and Astronomy, Rutgers University, Piscataway, New Jersey 08856-8019, USA} 
\date{\today} 
\begin{abstract}

We argue that, because of the quantum-entanglement,
the local physics of the strongly-correlated materials
at zero temperature is described
in very good approximation by a simple generalized Gibbs distribution,
which depends on a relatively small number local quantum thermodynamical potentials.
We demonstrate that our statement is exact in certain limits, and 
we perform numerical calculations of the iron compounds
FeSe and FeTe and of the elemental cerium by employing the Gutzwiller
Approximation (GA) that strongly support our theory in general.

\end{abstract}

\maketitle

The strongly-correlated electron-systems display an extremely rich variety of
phenomena, such as the Mott localization and the high-$T_{\text{c}}$ superconductivity,
which do not exist in conventional materials.
The key element at the basis of the unconventional physical effects exhibited
by the strongly-correlated materials is that
the Coulomb interaction ``localizes'' 
part of the electrons, which retain part of their atomic character
making it impossible to describe the system 
within a single-particle picture, 
and opening up the possibility of an
entirely different class of phenomena.

A fundamental object in order to understand the physics of
the strongly-correlated materials is the so-called
``reduced density-matrix'' of the correlated electrons, which is 
obtained from the exact density matrix of the solid 
by tracing over all degrees of freedom except for those of the 
correlated local orbitals of interest, e.g., the $d$-electrons of a transition-metal 
compound. 
In fact, this object encodes the whole local physics of the corresponding
electronic degrees of freedom. 
For instance, it enables us to study 
the average-populations, the
mixed-valence character~\cite{RevModPhys-mixvalence,deltaPu-gabi}
and the entanglement-entropy~\cite{RevModPhys-entropy,our-Ce} 
of the correlated orbitals, 
which are fundamental concepts
in modern condensed matter theory.

The scope of this work is to understand 
how the reduced density matrix of the correlated electrons 
is affected by the quantum environment in a solid at zero temperature.
Note that while this is a fundamental problem of great interest,
the answer is definitively non-trivial,
as the size of the reduced density matrix grows
exponentially with the number of correlated orbitals, and the interaction
between the local correlated orbitals and their environment
is generally very strong, and depends both on the chemical composition
and on the arrangement of the atoms within the solid.

Let us reformulate the problem 
from a general perspective, without confining explicitly the discussion
to the correlated electron-systems. 
We consider a generic ``large'' isolated system $\mathcal{U}$ 
(the lattice), and represent its Hamiltonian as
\be
\h_{\mathcal{U}} = \h_{\ms}+\h_{\mb}+\h_{\ms\mb}\,,\label{usb}
\ee
where $\h_\ms$ is the Hamiltonian of a subsystem $\ms$ (a subset of local atomic orbitals),
$\h_{\mathcal{B}}$ represents the Hamiltonian of its
environment $\mb$,  
and $\h_{\ms\mb}$  represents the interaction between $\ms$ and $\mb$.
Finally, we assume that $\mathcal{U}$ is in the ground-state 
$\ket{\Psi^{E_0}_\mU}$ of $\h_{\mathcal{U}}$ 
and we consider the corresponding reduced density-matrix
\be
\rs=\Tr_{\mathcal{B}}\,\ket{\Psi_\mU^{E_0}}\bra{\Psi^{E_0}_\mU}
\,.\label{rs}
\ee
How does $\rs$ depend on the coupling between $\ms$ and its environment?

In this work we argue that, because of the quantum-entanglement,
$\rs$ exhibits thermodynamical
properties pertinent to statistical averages.
More precisely we argue that, due to 
the property of $\ket{\Psi_\mU^{E_0}}$
to be quantum-entangled, $\rs$ has, approximately, a 
simple generalized Gibbs form, which depends only on few local
thermodynamical parameters.

Before to expose our theory it is useful to
discuss briefly an important recent related result: 
the canonical-typicality theorem~\cite{Goldstein,Popescu}.
This theorem states that, given 
a system represented as in Eq.~\eqref{usb}
--- with a \emph{very small} hybridization $\h_{\ms\mb}$, ---
the reduced density matrix $\rho_S$ of any
``typical'' $\ket{\Psi_\mU^E}\in\mU_{[E,E+dE]}$, 
where $\mU_{[E,E+dE]}$ is the Hilbert subspace generated by the eigenstates
of $\h_{\mathcal{U}}$ with energy in $[E,E+dE]$, is 
\be
\rs=e^{-\h_{\ms}/T_{\ms}}/\Tr[e^{-\h_{\ms}/T_{\ms}}]\,,\label{gibbs}
\ee
where the temperature $T_{\ms}$ is determined by the average energy 
$E_{\ms}\equiv\Tr[\hat{\rho}_S\h_{\ms}]$.
Note that the Gibbs form of $\rs$ arises as 
an \emph{individual} property
of the typical $\ket{\Psi_\mU^E}$
--- a \emph{pure} state, ---
without calling in cause the construction of an ensemble.
The key concept underlying this important theorem 
is the quantum-entanglement. 
A simple way to make this interpretation clear is 
that Eq.~\eqref{gibbs} is characterized by the condition 
\be
S[\rs]
=\max \{S[\hat{\rho}]\,|\;\Tr[\hat{\rho}\h_{\ms}]=E_{\ms}\}\label{sg}
\,,
\ee
where $S[\hat{\rho}]=-\Tr[\hat{\rho}\log\hat{\rho}]$
is the EE of $\ms$, and $\Omega_{\ms}$ is the set of all 
of the local (that is, in $\ms$) density-matrices.
Since the 
entanglement-entropy is a measure of the quantum-entanglement between
$\ms$ and $\mb$, this characterization 
of $\rs$ shall be regarded as a consequence of the 
individual property of the typical $\ket{\Psi_\mU^E}\in\mU_{[E,E+dE]}$
to be highly-entangled~\cite{Eisert}.

Let us now drop the assumption that the interaction
between $\ms$ and $\mb$ is negligible (which is certainly not the case in materials),
and focus 
on our questions concerning 
the reduced density matrix $\rs$ of the
ground-state $\ket{\Psi^{E_0}_\mU}$ of $\mU$, see Eq.~\eqref{rs} and text below.
The key message of this work is that, even in this case, 
a proper generalization of Eq.~\eqref{gibbs} holds --- albeit only
approximately.

In order to demonstrate our statement, let us consider 
the density-matrix $\hat{\rho}(a_1,..,a_n)$ characterized by the condition
\be
S[\hat{\rho}(a_1,..,a_n)]
=\max \{S[\hat{\rho}]\,|\;\Tr[\hat{\rho}\hat{A}_i]=a_i\;\forall i\}\label{sg-gen}
\,.
\ee
It is known that
the solution $\hat{\rho}(a_1,..,a_n)$ of Eq.~\eqref{sg-gen} 
has, if it is nondegenerate,
the following ``generalized canonical'' form~\cite{cme1,cme2,cme3}
\be
\hat{\rho}(\lambda_1,..,\lambda_n)
=e^{-\sum_{i=1}^n \lambda_i\hat{A}_i}/\Tr[e^{-\sum_{i=1}^n \lambda_i\hat{A}_i}]\,.
\label{cmeea}
\ee
From now on we refer to Eq.~\eqref{cmeea} as the
\emph{principle of maximum entanglement entropy} (PMEE) 
relative to the set of observables $\{\hat{A}_1,..,\hat{A}_n\}$,
and to the constraints in Eq.~\eqref{sg-gen}
as the corresponding ``testable information''.

A possible way to quantify the goodness of a given PMEE 
is the following quantity:
\be
\Delta[\{\hat{A}_1,..,\hat{A}_n\}]
=\min_{\lambda_1,..,\lambda_n} D[\hat{\rho}(\lambda_1,..,\lambda_n),\rs]
\,,\label{measure-leg}
\ee
where $\rs$ is the actual reduced density matrix of the system, and
\be 
D[\hat{\rho}_1,\hat{\rho}_2]\equiv 
\Tr\!\left(|\hat{\rho}_1-\hat{\rho}_2|\right)/2
\in [0,1]\label{distance}
\ee
is a standard trace-distance~\cite{book-quantum-information}, 
that represents
the maximal difference between $\hat{\rho}_1$ and $\hat{\rho}_2$ in the probability
of obtaining any measurement outcome.

In summary, we have proposed 
a systematic method to construct and 
verify the goodness of a generalized Gibbs ansatz
for the reduced density-matrix $\rs$ of a generic system.
The key step is the identification of a 
subset of local observables 
$\hat{\mathcal{A}}\equiv\{\hat{A}_1,..,\hat{A}_n\}$,
whose expectation-values are expected ---
e.g., on the basis of physical considerations
--- to be directly ``controlled'' by the system-environment interaction.
Note that $\Delta[\hat{\mathcal{A}}]\equiv 0$
in the limit in which $\hat{\mathcal{A}}$ 
coincides with the set of all of the local observables.
In fact, any density-matrix is uniquely defined by all of the
expectation-values of the observables within its Hilbert space.

As we are going to show, the PMEE
is a very useful theoretical tool, as a subset $\hat{\mathcal{A}}$
containing only ``few'' observables is often sufficient
to have $\Delta[\hat{\mathcal{A}}]\simeq 0$.
In other words, it is generally possible to define
a series of observables $\hat{A}_i$ such that the corresponding series
of trace-distances
\be
\Delta_n\equiv\Delta[\{\hat{A}_1,..,\hat{A}_n\}]
\label{dn}
\ee 
converges ``rapidly'' to zero as a function of $n$,
regardless the details of the environment $\mb$ and
its coupling with $\ms$.

We point out that the PMEE 
[Eq.~\eqref{cmeea}]
has a twofold interpretation:
(1) the only ``relevant'' testable information of $\rs$ consists in
the expectation-values $a_i$ of the observables $\hat{A}_i\in\hat{\mathcal{A}}$;
(2) the $\ms$ degrees of freedom 
are essentially in a Gibbs state, but they experience 
the effective interaction encoded in a ``renormalized'' local Hamiltonian 
\be
\hat{\F}_\ms=\sum_{i=1}^n \lambda_i\hat{A}_i
\,\quad (\hat{A}_i\in\hat{\mathcal{A}})\,,
\label{fdef}
\ee
that is generally different from 
to the original $\h_{\ms}$.
This twofold interpretation reflects the 
Legendre-duality between the expectation-values $a_i$ and the corresponding 
generalized chemical potentials $\lambda_i$.

\emph{Strongly-correlated electron-systems.}----
From now on we restrict the attention to the many-body correlated electron-systems
in their ground state. 
More precisely, we consider a generic multi-band Hubbard model (HM)
\be
\h_{\mathcal{U}}=\sum_{i\neq j}\sum_{a,b=1}^{\nu}\epsilon^{ab}_{ij}\, \cc_{ia}\ca_{jb}
+\sum_i\hl_i[\{\cc_{ic}\},\{\ca_{ic}\}]\,;\label{hubbard}
\ee
where $i,j$ are ``site'' labels and $a,b,c=1,..,\nu$ label both
the spin $\sigma$ and the orbital $m$. 
The Hamiltonian $\h_{\mathcal{U}}$ can be separated as in Eq.~\eqref{usb}, 
with $\h_{\ms}$ corresponding to the $i$-local operator $\hl_{i}$ ---
which, in general, can include
both a quadratic term $\h^{\epsilon}_i$ and a quartic term
$\h^{\text{int}}_i$ (representing the on-site Coulomb interaction).

In order to define a PMEE 
for the $\ms$ reduced density-matrix we need to understand
which local observables have to be included in $\hat{\mathcal{A}}$,
see Eq.~\eqref{fdef},
to describe approximately the local physics of the system.

Due to the coupling between the environment and the local space,
the expectation-value of $\h_{\ms}$ with respect to $\rs$
is controlled by their reciprocal interaction.
In other words, $\ms$ and $\mb$ exchange energy (as in the canonical case),
which implies that the expectation-value of 
$\h_{\ms}$ has to be included in $\hat{\mathcal{A}}$.
On the other hand, since $\h_{\ms\mb}$ is \emph{not} generally small,
there are at least two additional key physical mechanisms that our PMEE 
shall take into account.
(I) Due to the hybridization effect, 
also the individual local orbital populations 
are controlled by the coupling with the environment.
(II) We expect that 
the effective local interaction
$\hat{\F}_\ms$ experienced by the local degrees of freedom is renormalized.
Furthermore, we expect that $\hat{\F}_\ms$ is 
not isotropic, but is 
invariant only under the point-group of the system.

From the above heuristic arguments, we conclude 
that $\hat{\mathcal{A}}$
should include \emph{at least}
all of the quadratic and quartic operators compatible with the 
symmetry of the system.
According to our scheme, the corresponding PMEE is 
\be
\hat{\rho}_\ms^{\text{fit}} 
=e^{-\hat{\F}_\ms}/\Tr[e^{-\hat{\F}_\ms}] 
\,,\label{ansatz1}
\ee
where the operator $\hat{\F}_\ms$ is a generic 
linear combination of quadratic
and quartic operators.
Note that Eq.~\eqref{ansatz1} represents
an extremely ``special'' density-matrix, since the number of parameters that determine
it grows only \emph{quartically} with $\nu$,
see Eq.~\eqref{hubbard},
rather than \emph{exponentially}.

We point out that Eq.~\eqref{ansatz1} is 
exact not only in the so called ``atomic limit'' 
$\h_{\ms\mb}\rightarrow 0$, but
also for any quadratic $\h_\mU$~\cite{free-red-mat}.
This is a simple consequence of the Wick's theorem, which ensures
that the expectation-value of any local observable
depends only on the ``Wick's contractions''
$\Av{\Psi_{\mathcal{U}}^{E_0}}{\cc_{\ms a}\ca_{\ms b}}$,
that can be readily reproduced by Eq.~\eqref{ansatz1} within
a proper quadratic $\hat{\F}_\ms$.

For later convenience, we define the following series of PMEE for
the local reduced density matrix $\rs$ of a generic HM.
(i) $\Delta_1$, corresponding to $\hat{\mathcal{A}}_1\equiv\{\h^{\text{loc}},\hat{N}\}$,
where $\h^{\text{loc}}$
is the on-site Hamiltonian and $\hat{N}$ is
number operator of $\ms$-electrons;
(ii) $\Delta_2$, corresponding to $\hat{\mathcal{A}}_2$ containing 
$\h^{\text{loc}}$ and all of
the quadratic operators commuting with the point group of the system;
(iii) and $\Delta_3$, corresponding to $\hat{\mathcal{A}}_3$ containing
all of the quadratic and the quartic operators
that commute with the point group of the system.
As a reference, it is also useful to define the 
trace-distance $\Delta_0$ between $\rs$
and the maximally-entangled state, which is the local
density-matrix proportional to the identity
--- that corresponds to the PMEE for an empty set of observables,
$\hat{\mathcal{A}}_0\equiv\{\}$.

\emph{The iron chalcogenides.}----
In order to benchmark our theoretical arguments
and demonstrate their utility to the study of materials,
here we consider, as a first example,
the reduced local density matrix $\rho_d$
of a realistic HM
representing the iron compound FeSe.

We construct the HM of FeSe adopting 
the same bands-structure $\epsilon$, see Eq.~\eqref{hubbard}, 
used in Ref.~\cite{our-FeSe-FeTe},
which was generated 
using Density Functional Theory with the Generalized Gradient
Approximation for the exchange-correlation potential, according to
the Perdew-Burke-Ernzerhof recipe implemented in
Quantum Espresso [26], and applying Wannier90 [27] to compute
the maximally localized Wannier orbitals.
Finally, we make use of the Slater parametrization of the on-site interaction $\h^{\text{int}}$.
Since the HM can not be solved exactly, we solve it approximately
within the GA~\cite{Gutzwiller3}, which is 
a very reliable approximation for the ground-state of 
the correlated metals.
In particular, we employ the numerical implementation developed
in Refs.~\cite{Deng_LDA+Gutz,Gmethod,full-slater,DMFTG}.

\begin{figure}
\begin{center}
\includegraphics[width=8.6cm]{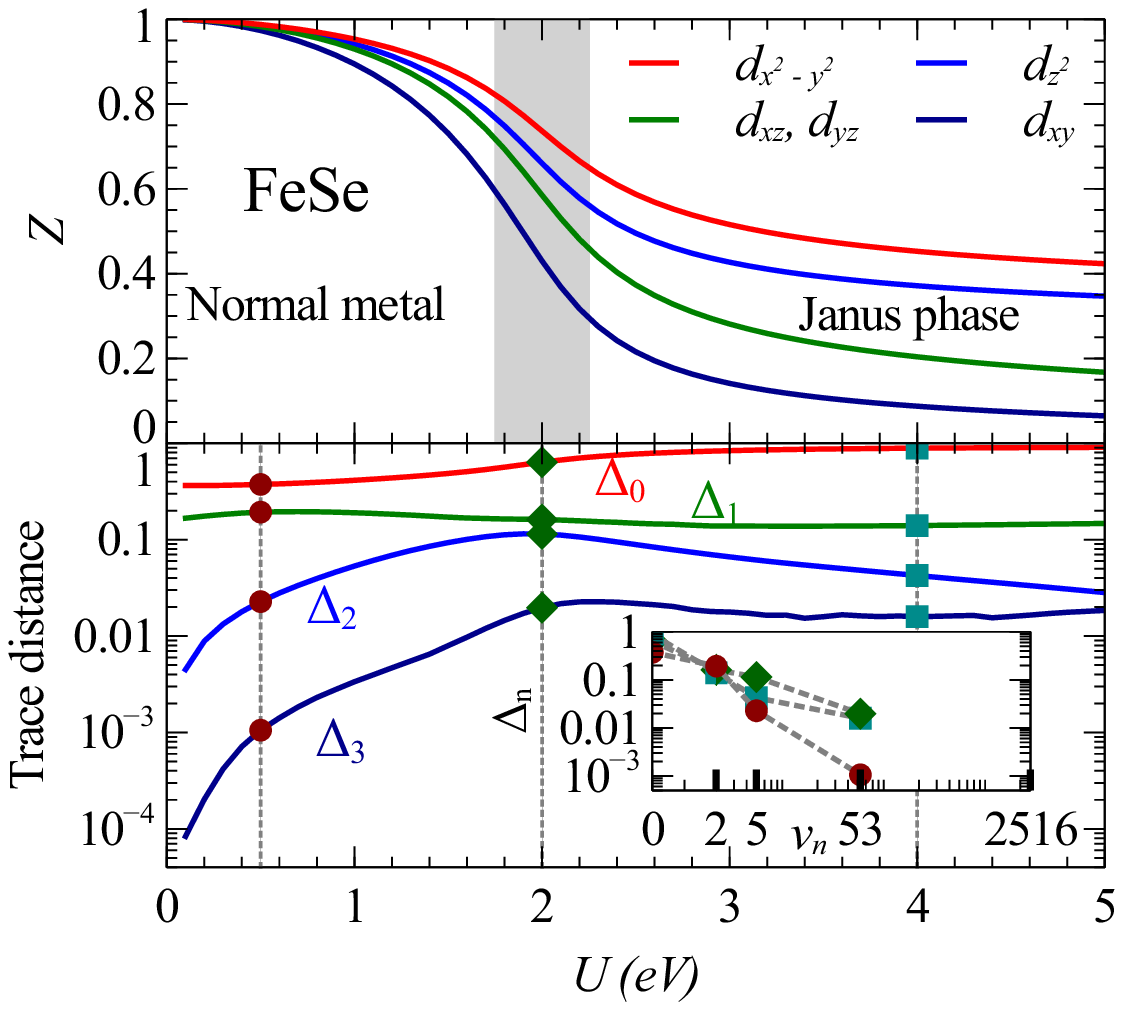}
\caption{(Color online) Upper panel: quasi-particle renormalization weights of FeSe.
  The normal-metal phase (small $U$) and the Janus phase (large $U$) 
  are indicatively separated by a vertical shaded line.
  Lower panel: PMEE trace-distances $\Delta_n$ 
  for the reduced density matrix $\rho_d$ of FeSe.
  The series shown in the insets correspond to the three values of $U$
  indicated by vertical lines in the main panel.
  The calculations are performed at fixed $J/U=0.224$ as a function of $U$.
}
\label{figure1}
\end{center}
\end{figure}

In the first panel of Fig.~\ref{figure1} the quasi-particle renormalization 
weights are shown as a function of the
interaction-strength $U$, keeping the ratio $J/U$ fixed at $0.224$.
As discussed in Ref.~\cite{our-FeSe-FeTe}, at $U\simeq 2\,eV$ the system
undergoes a clear crossover from a normal metallic phase ($Z\simeq 1$) toward a bad-metallic
phase ($Z\ll 1$) --- the so called Janus phase~\cite{Janus}.
Our purpose is to
analyze the local reduced density-matrix $\hat{\rho}_d$ of the Fe $d$ electrons
and to verify the goodness of the PMEE for our FeSe HM,
both in its normal-metal regime and in its Janus phase.

In the second panel of Fig.~\ref{figure1} is shown
the evolution of the 
PMEE trace-distances $\Delta_0,\Delta_1,\Delta_2,\Delta_3$.
The corresponding series
$\Delta_n$, see Eq.~\eqref{dn},
is shown explicitly in the inset for three values of $U$
as a function of the respective number $\nu_n$ of fitting parameters required.
Remarkably, $\Delta_n$ converges very rapidly to $0$ for all
$U$'s, see Eq.~\eqref{dn}.
In fact, although the number of independent parameters of $\hat{\rho}_d$ is $2516$,
the $\Delta_3$ PMEE, which is defined by only 
$53$ free parameters, is sufficient
to obtain a very accurate fit for every $U$ considered
--- as indicated by the trace-distance $\Delta_3\ll 1$.

\begin{figure}
\begin{center}
\includegraphics[width=8.6cm]{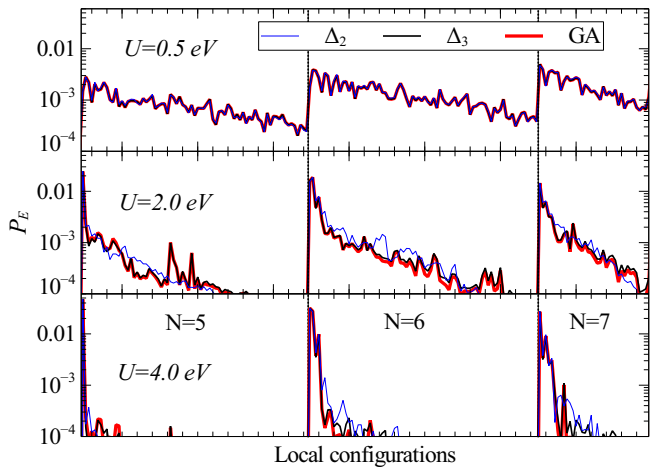}
\caption{(Color online) Histogram of local configuration probabilities
  $P_E$ of the eigenstates of $\h^{\text{loc}}$ of FeSe,
  in the sectors $N=5,6,7$,
  for $U=0.5\,eV$, $U=2\,eV$ and $U=4\,eV$ at fixed $J/U=0.224$. 
  The GA configuration probabilities (red line) are shown 
  in comparison with the local configuration probabilities 
  evaluated using the PMEE density matrices corresponding to the distances
  $\Delta_2$ and $\Delta_3$.
  Within each $N$-sector, the configuration probabilities $P_E$ are sorted
  in ascending order of 
  energy $E\equiv\Av{\psi_E}{\h^{\text{loc}}}$, where $\ket{\psi_E}$ are the
  eigenstates of $\h^{\text{loc}}$.
}
\label{figure2}
\end{center}
\end{figure}

In order to get an even better idea of how accurate our PMEE fits are,
we show also the histogram of the
local configuration probabilities of the eigenstates of $\h^{\text{loc}}$,
\be
P_E\equiv\Tr\!\left[\hat{\rho}_d\,\hat{P}_E\right]/d_E\,,
\ee
where $\hat{P}_E$ is the orthonormal projector over the $E$-eigenspace of $\h^{\text{loc}}$ and
$d_E$ is its degeneracy. 
In Fig.~\ref{figure2}
the computed GA configuration probabilities are shown 
for three values of $U$
in comparison with the local configuration probabilities 
evaluated using the PMEE density matrices corresponding to the trace-distances
$\Delta_2$ and $\Delta_3$.
These data confirm the goodness of our PMEE.
In fact, the structure of the computed local
configuration probabilities, which is extremely complex,
is captured in detail by the $\Delta_3$ PMEE, and
this agreement is verified
for all $U$'s, even though the system undergoes a clear crossover between two
electronically distinct phases at $U\simeq 2$.

The same calculations shown above for FeSe are also reported for FeTe
in the supplementary material. 
The results are essentially identical, which further
supports our theory.


\emph{The $\gamma$-$\alpha$ transition of cerium.}---- 
As a second example, we consider the elemental cerium at zero
temperature, which has been recently studied theoretically within
the charge self-consistent
Local Density Approximation in combination with the GA,
see Ref.~\cite{our-Ce}.
In Ref.~\cite{our-Ce}, the important role of the spin-orbit
coupling for the $\gamma$-$\alpha$ iso-structure transition 
has been understood by observing the rapid variation of the 
EE of the $f$ electrons,
in correspondence of the
signature of the $\gamma$-$\alpha$ transition --- i.e., concomitantly to the
minimum of the bulk-modulus $\mathcal{K}=-VdP/dV$.

Let us consider the
$\Delta_2$ PMEE for the $f$ local reduced density matrix $\hat{\rho}_f$,
\be
\hat{\rho}_f^{\text{fit}}\propto 
e^{-\left[\hl+\left(\delta\mu_{5/2}\,\hat{n}_{5/2}+\delta\mu_{7/2}\,\hat{n}_{7/2}\right)\right]/\tau}\,.
\label{Ce-ansatz}
\ee
Equation~\eqref{Ce-ansatz} corresponds to maximize the $f$-EE at given average number of
$5/2$ and $7/2$ $f$-electrons and given expectation-value of $\hl\equiv\h^{\text{int}}+\h^{\text{soc}}$;
where 
$\h^{\text{int}}$ is the Slater interaction with $U=6\,eV$ and $J=0.7\,eV$,
and $\h^{\text{soc}}$ is the spin-orbit-coupling (SOC) operator
\be
\h^{\text{soc}}\equiv\mu_{5/2}\,\hat{n}_{5/2}+\mu_{7/2}\,\hat{n}_{7/2}\,,
\ee
where the coefficients $\mu_{5/2}$ and $\mu_{7/2}$ are obtained from the LDA 
Kohn-Sham Hamiltonian.

Note that the parameter $\tau$ of Eq.~\eqref{Ce-ansatz}
is \emph{not} the physical temperature 
--- which is zero by definition in our calculations.
In fact, the expectation-value of $\hl$,
that is the Legendre-conjugate variable of $\beta_\tau\equiv 1/\tau$,
is necessarily always higher than its ground-state energy, as the cerium atoms
are not isolated, but they are embedded in the fcc lattice-structure.
The parameters 
$\delta\mu_{5/2}$ and $\delta\mu_{7/2}$ renormalize
the bare spin-orbit splitting ``experienced'' by the $f$ electrons,
which is 
\be
\delta_{\text{soc}}\equiv
\left(\mu_{7/2}+\delta\mu_{7/2}\right)-\left(\mu_{5/2}+\delta\mu_{5/2}\right)\,.
\label{deltasoc}
\ee

\begin{figure}
\begin{center}
\includegraphics[width=8.6cm]{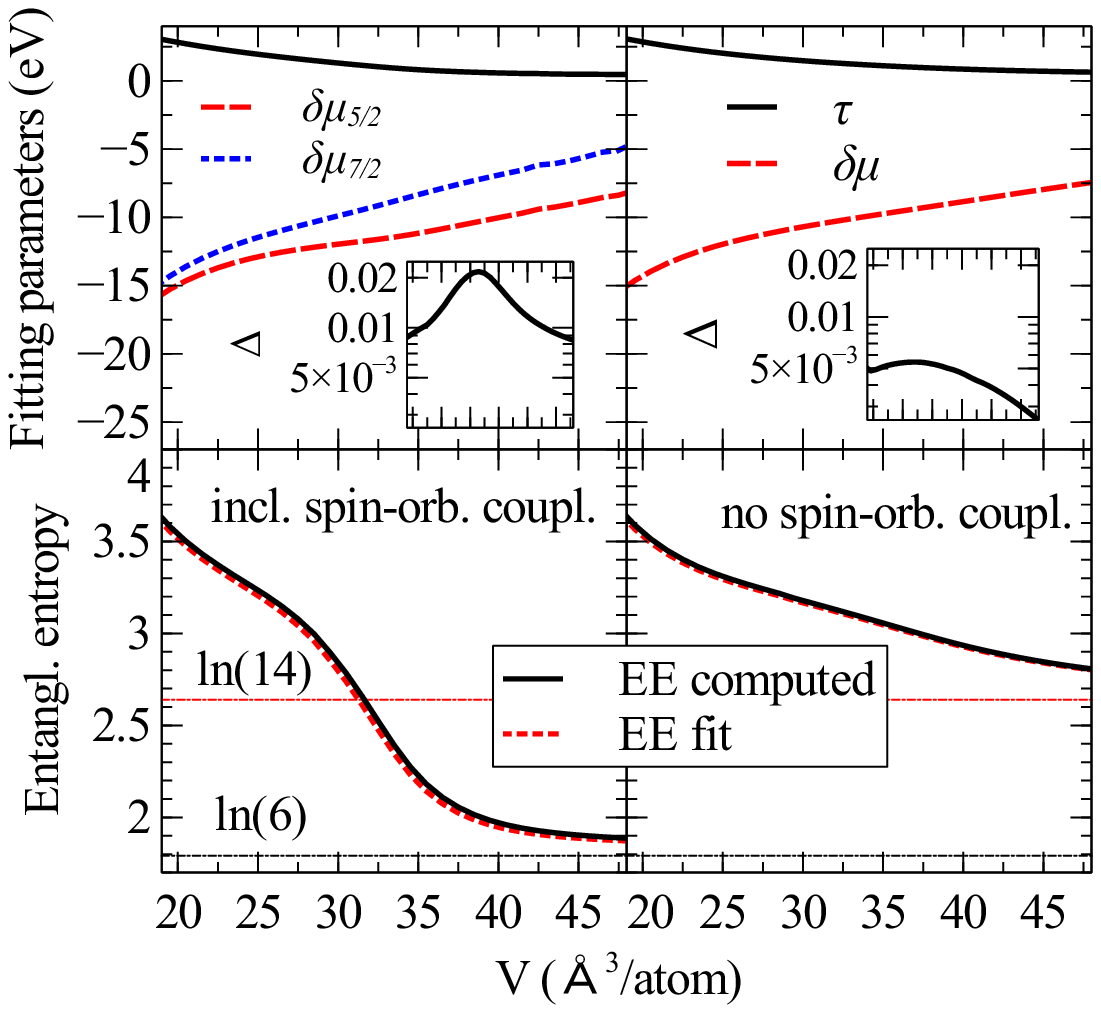}
\caption{(Color online)
Evolution of the PMEE fitting parameters as a function of the volume (upper panels)
and corresponding trace-distances (insets).
In the lower panel the evolutions of the EE of the
computed and fitted reduced $f$ density matrices are also shown.
Our results are shown both with (left panels) 
and without (right panels) taking into account the spin-orbit coupling.
}
\label{figure3}
\end{center}
\end{figure}

In the upper panel of Fig.~\ref{figure3} 
is shown the behavior 
as a function of the volume of $\Delta_2$
and of the fitting parameters
$\tau$, $\delta\mu_{5/2}$ and $\delta\mu_{7/2}$.
In the lower panel also the evolution of $S[\hat{\rho}_f]$ 
and $S[\hat{\rho}_f^{\text{fit}}]$
is illustrated. 
As in Ref.~\cite{our-Ce},
our results are shown both by taking into account the SOC and by neglecting it.

We point out that the PMEE 
$\hat{\rho}_f^{\text{fit}}$, see Eq.~\eqref{Ce-ansatz},
approximates very well the computed reduced density matrix $\hat{\rho}_f$,
as indicated by the trace-distance $\Delta_2\ll 1$.
The effective temperature $\tau$ decreases by increasing the
volume. This is to be expected, as 
$\tau$ reduces to the actual physical temperature $T=0$ in the infinite-volume limit. 
On the contrary, at large volumes the renormalized spin-orbit splitting
$\delta_{\text{soc}}$
becomes considerably larger than $\mu_{7/2}-\mu_{5/2}$, 
which is of the order of $0.3\,eV$ for all volumes considered (not shown).

Remarkably, although the PMEE 
fitting parameters vary \emph{smoothly}
for all volumes, the entanglement entropy 
$S[\hat{\rho}_f]$ changes \emph{rapidly} at 
$V\simeq 33\,\AA/\text{atom}$ --- i.e., 
around the minimum of the bulk-modulus $\mathcal{K}$~\cite{our-Ce}, --- 
indicating that $\hat{\rho}_f$ undergoes a rapid crossover.
This observation enables us to ascribe the $\gamma$-$\alpha$ transition
of cerium  
to the (approximate) generalized Gibbs form [Eq.~\eqref{Ce-ansatz}] of $\hat{\rho}_f$,
and to describe the role of the spin-orbit coupling
very neatly as follows.
Analogously  to a two-level system with a single particle in thermal equilibrium 
--- in which the equilibrium state undergoes a crossover when the temperature
becomes comparable with the energy-gap between the two levels, ---
the 
crossover of the
reduced density matrix $\hat{\rho}_f$
relates with the \emph{discrete} structure of the spectrum of 
$\hat{\mathcal{F}}^{\text{loc}}\equiv\hl+\left(\delta\mu_{5/2}\,\hat{n}_{5/2}+\delta\mu_{7/2}\,\hat{n}_{7/2}\right)$
in relation with the fictitious temperature $\tau$.

The physical picture outlined above
might be applicable to describe the volume-collapse transitions in $4f$
and $5f$ systems in general,
which is a major subject of investigation in condensed matter physics.



\emph{Conclusions.}----
We have shown that the local physics of the strongly-correlated materials
at zero temperature is described by a simple universal generalized
Gibbs distribution. This statement is deeply significant, as
the interaction between the subsystem (a given atom) 
and its environment (all of the other atoms) is 
definitively non-negligible in real materials.
Our finding provides a very powerful theoretical viewpoint on the
strongly-correlated electron-systems.
In fact, as shown explicitly by our calculations, 
the simple exponential form of the reduced density matrix 
enables us to understand 
in terms of few local thermodynamical parameters
the behavior of many important physical quantities, such as all of 
the many-body local configuration probabilities of the correlated electrons
--- whose number is extremely large in general, as it grows exponentially
with the number of correlated orbitals.
In the future, our finding might open up the possibility
to engineer new compounds with desired physical local properties
by directly controlling the local thermodynamical parameters, e.g.,
through proper structure modifications.
Finally, our general interpretation of this results, that is
based on the quantum entanglement, suggests
that our theory might be applicable not only to materials, but also to other 
quantum systems.

\begin{acknowledgments}
  We thank Sheldon Goldstein, Xiaoyu Deng, 
  Giovanni Morchio, Michele Fabrizio,
  Cai-Zhuang Wang and Kai-Ming Ho
  for useful discussions.
  The collaboration was supported by the U.S. Department of Energy
  through the Computational Materials and Chemical Sciences Network CMSCN.
  Research at Ames Laboratory supported by the U.S.
  Department of Energy, Office of Basic Energy Sciences,
  Division of Materials Sciences and Engineering. 
  Ames Laboratory is operated for the U.S. Department of Energy
  by Iowa State University under Contract No. DE-AC02-07CH11358.
  H. U. R. S. acknowledges the support of the Mathematics - Physics Platform
  ($\mathcal{MP}^{\textsf{2}}$) at the University of Gothenburg. Simulations
  were performed on resources provided by the Swedish National Infrastructure
  for Computing (SNIC) at Chalmers Centre for Computational Science and
  Engineering (C3SE) (project no.~01-11-297).
\end{acknowledgments}

\bibliographystyle{apsrev}


\newpage

\section*{SUPPLEMENTAL MATERIAL}

\subsection*{PMEE results for FeTe}

\begin{figure}[H]
\begin{center}
\includegraphics[width=8.6cm]{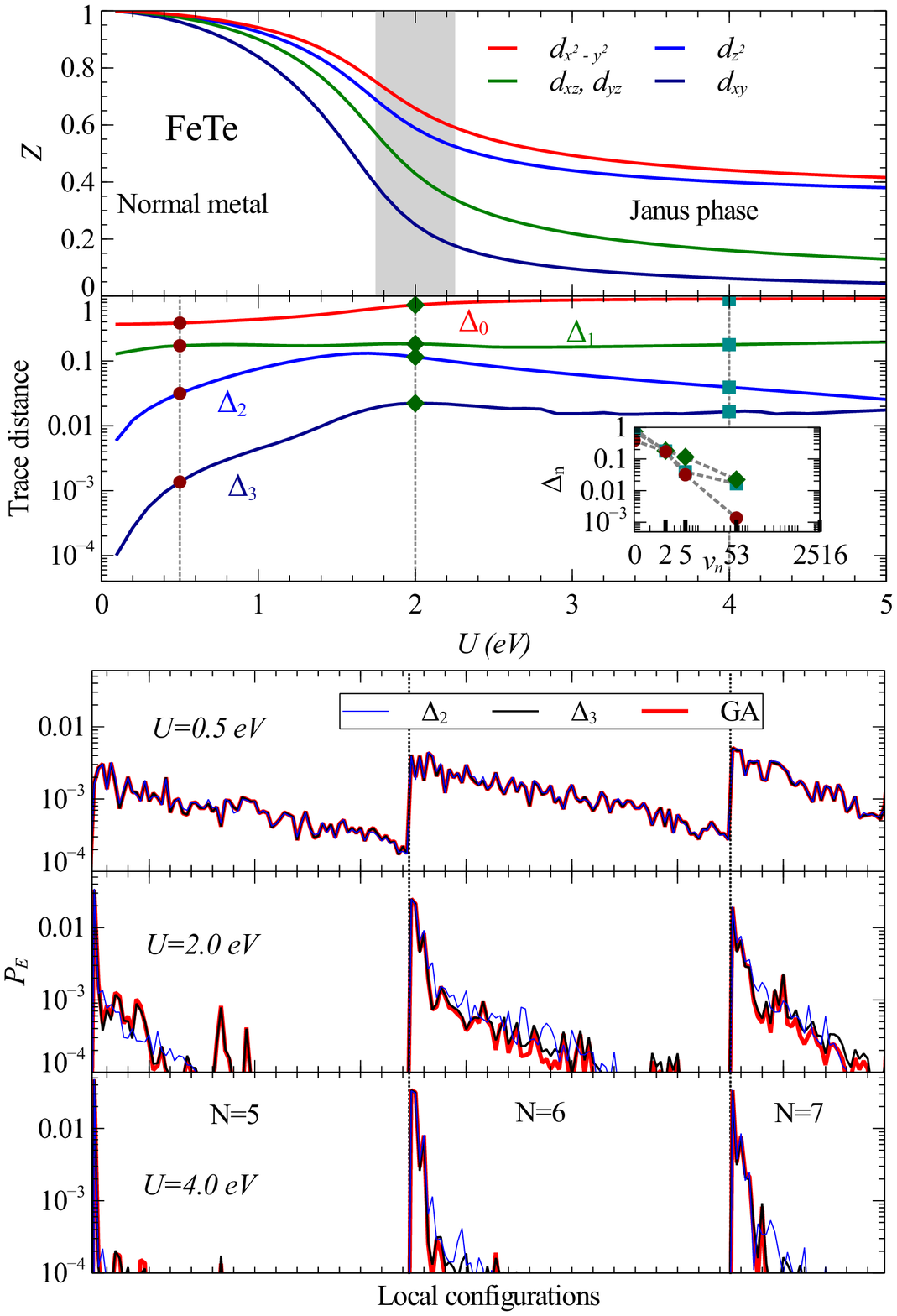}
\caption{(Color online) First panel: quasi-particle renormalization weights of FeTe.
  The normal-metal phase (small $U$) and the Janus phase (large $U$) 
  are indicatively separated by a vertical shaded line.
  Second panel: PMEE trace-distances $\Delta_n$ 
  for the reduced density matrix $\rho_d$ of FeSe.
  The series shown in the insets correspond to the three values of $U$
  indicated by vertical lines in the main panel.
  The calculations are performed at fixed $J/U=0.224$ as a function of $U$.
  \\
  Lower panels: Histogram of local configuration probabilities
  $P_E$ of the eigenstates of $\h^{\text{loc}}$ of FeTe,
  in the sectors $N=5,6,7$,
  for $U=0.5\,eV$, $U=2\,eV$ and $U=4\,eV$ at fixed $J/U=0.224$. 
  The GA configuration probabilities (red line) are shown 
  in comparison with the local configuration probabilities 
  evaluated using the PMEE density matrices corresponding to the distances
  $\Delta_2$ and $\Delta_3$.
  Within each $N$-sector, the configuration probabilities $P_E$ are sorted
  in ascending order of 
  energy $E\equiv\Av{\psi_E}{\h^{\text{loc}}}$, where $\ket{\psi_E}$ are the
  eigenstates of $\h^{\text{loc}}$.
}
\label{sup}
\end{center}
\end{figure}

\end{document}